\documentstyle[pre,aps]{revtex}

\begin{document}

\draft

\title{Nonground State Condensates of Ultracold Trapped Atoms}

\author{V.I. Yukalov$^{1,2}$, E.P. Yukalova$^{1,3}$ and V.S. Bagnato$^1$}

\address{$^1$Instituto de Fisica de S\~ao Carlos, Universidade de S\~ao Paulo \\
Caixa Postal 369, S\~ao Carlos, S\~ao Paulo 13560-970, Brazil} 
\address{$^2$Bogolubov Laboratory of Theoretical Physics\\
Joint Institute for Nuclear Research, Dubna 141980, Russia} 
\address{$^3$Laboratory of Computing Techniques and Automation\\
Joint Institute for Nuclear Research, Dubna 141980, Russia}

\maketitle

\begin{abstract}

The population dynamics of a trapped Bose-Einstein condensate, subject 
to the action of an oscillatory field, is studied. This field produces
a modulation of the trapping potential with the frequency close to the
transition frequency between the ground state and an excited energy level.
Unusual critical effects are found exhibiting sharp qualitative changes 
in the population dynamics. An effective averaged system is constructed 
explicitly illustrating the occurrence of critical phenomena. The related 
critical indices are calculated.

\end{abstract}

\vskip 1cm


Bose-Einstein condensates of trapped atoms present an interesting example of
essentially nonlinear statistical systems with rich properties that are being
intensively studied both experimentally and theoretically [1,2]. Although
experiments deal with rather dilute Bose gases, their interaction,
nevertheless, cannot be considered as a small perturbation. This is because
in a system of condensed atoms coherence develops, resulting in an
effective interaction proportional to the number of
particles. The fact that Bose-Einstein condensation and coherence appear
simultaneously can be naturally understood remembering that both these
phenomena arise when the thermal wavelength of atoms exceeds the mean
interatomic distance. All the measurements done so far show the evidence
for coherence of Bose-Einstein condensates [1,2]. Essential nonlinearity of
the latter makes their behaviour quite different from that of ensembles
of weakly interacting gases.

In the present article we show that the nonlinearity in the Hamiltonian
describing  Bose condensates results in very unusual dynamical effects
resembling critical phenomena in statistical systems. We consider the
dynamics of population levels of a Bose condensate subject to the action
of a resonant field whose frequency is close to the transition frequency
between the ground-state and a chosen excited level. Varying the
characteristic parameters, such as the transition amplitude and detuning,
we show that there exists a bifurcation line, dividing the space of these
parameters onto regions, where the dynamical behaviour of the system is
dramatically different. This bifurcation line is analogous to the critical
line of a statistical system experiencing critical phenomena. To substantiate
this analogy, we construct an effective system explicitly demonstrating
critical behaviour along the critical line coinciding with the bifurcation
line found for the evolution equations. For the effective system, it is
possible to define the quantities analogous to the order parameter, specific
heat, and susceptibility and even to calculate their critical indices.

Let us consider a Bose system, when all atoms are condensed and the system
can be described by the nonlinear Schr\"odinger equation, often called the
Gross-Pitaevskii equation [1,2]. Assume that, in addition to a stationary
trapping potential, there is a time-dependent potential, which we
call the resonance field and whose meaning will be specified below.
Thus, we consider the equation
\begin{equation}
\label{1}
i\hbar\;\frac{\partial\varphi}{\partial t} = \left [ \hat H(\varphi) +
V_{res}\right ]\; \varphi\; , \qquad
\hat H(\varphi)= \; - \; \frac{\hbar^2}{2m_0}\; {\vec\nabla}^2 +
U(\vec r\;) + A|\varphi|^2 \; ,
\end{equation}
with the nonlinear Hamiltonian in which $U$ is a trapping potential; 
$A\equiv4\pi\hbar^2a_sN/m_0$; $a_s$ is a scattering length; $m_0$ is 
mass; and $N$ is the number of atoms. The resonance field has the form
$V_{res} =V(\vec r\;)\cos\omega t$. This can be treated as the modulation 
of the trapping potential.

Tuning the frequency $\omega$ of the resonant field close to the
transition frequency $\omega_{mn}\equiv(E_m-E_n)/\hbar$ between the energy
levels defined by the stationary problem $\hat H(\varphi_n)\varphi_n=
E_n\varphi_n$, we may realize the corresponding interlevel transitions [3].
Since the nonlinear Schr\"odinger equation is an exact equation for the wave
function of a coherent state [4], the stationary solutions $\varphi_n$
can be called {\it coherent modes}. These should not be confused with
collective excitations defined as small deviations from the ground-state
and described by a linearized equation [1,2], although resonant 
transitions between different branches of collective excitations are also
possible [5]. But the equation for coherent modes is principally nonlinear.

Suppose that at the initial time $t=0$ the Bose gas was condensed, with 
all atoms being in the ground state $\varphi(\vec r,0)=\varphi_0(\vec r\;)$.
After this, the resonance field is switched on, with a frequency $\omega$ 
being in quasiresonance with the transition frequency $\omega_{j0}$ between 
the ground-state level and a chosen energy level  $j$. The quasiresonance 
condition is $|\Delta\omega|\ll\omega_{j0}$, where the detuning 
$\Delta\omega\equiv\omega-\omega_{j0}$. The solution of Eq. (1) can be 
presented as a sum $\varphi(\vec r,t) = \sum_n\; c_n(t)\;\varphi_n(\vec r,t)$ 
over coherent modes $\varphi_n(\vec r,t)=\varphi_n(\vec r)\exp(-iE_nt/\hbar)$.
Substituting this expansion in Eq. (1), one gets a system of equations 
for the coefficients $c_n(t)$. These equations can be simplified employing 
the quasiresonance condition. To this end, it is convenient to introduce 
the notation for the fractional level population $n_i(t) \equiv |c_i(t)|^2$,
and for the interaction intensity $\alpha$ and the transition amplitude
$\beta$, respectively,
\begin{equation}
\label{2}
\alpha\equiv\; \frac{A}{\hbar}\; \int\; |\varphi_0(\vec r\;)|^2
(2|\varphi_j(\vec r\;)|^2-|\varphi_0(\vec r)|^2)\; d\vec r\; , \qquad
\beta\equiv \; \frac{1}{\hbar}\; \int\; \varphi_0^*(\vec r\;) V(\vec r\;)
\varphi_j(\vec r\;)\; d\vec r\; .
\end{equation}
Then, in the quasiresonance approximation [3], we have the system of 
equations 
\begin{equation}
\label{3}
\frac{dc_0}{dt} =\; - i\alpha\; n_jc_0 -\; \frac{i}{2}\; \beta\;
e^{i\Delta\omega t} c_j\; , \qquad
\frac{dc_j}{dt} =\; - i\alpha\; n_0c_j -\; \frac{i}{2}\; \beta^*\;
e^{-i\Delta\omega t} c_0
\end{equation}
for the dynamics of the ground-state coefficient $c_0(t)$ and of the
coefficient $c_j(t)$ related to a chosen excited mode $j$. According to 
the assumption that at the initial time all atoms are condensed in the 
ground state $\varphi_0(\vec r\;)$, the initial conditions to Eqs. (3) are
$c_0(0)=1$ and $c_j(0)=0$. The derivation of Eqs. (3) has been explained 
in detail in Ref. [3]. However, it is worth emphasizing here a couple of 
important points. Looking for a solution of Eq. (1) in the form of the 
sum $\sum_nc_n\varphi_n$ does not require that the modes $\varphi_n$ 
compose a  complete set. Recall that the completeness of a basis is the 
possibility  of expanding over it an {\it arbitrary} function from the 
considered space. Such an excessively restrictive property is of no need 
for us. What we need is just to present the {\it sole} function $\varphi$ 
as a sum over modes, with the coefficients $c_n$ to be defined from Eq. (1). 
In the  theory of differential equations, this way of looking for a solution 
is called the method of substitution or the method of variation of 
parameters. This method is often used for solving nonlinear differential 
equations, e.g. for nonlinear optical waveguide equations [6,7] and for 
the nonlinear Schr\"odinger equation [8,9]. An exact orthogonality of the 
nonlinear modes is also not necessary, but, in general, it is sufficient 
that the modes $\varphi_n$ be approximately orthogonal in the sense of the 
smallness of the scalar product $(\varphi_m,\varphi_n)$ for $m\neq n$.
In the case of the nonlinear Schr\"odinger equation, taking for coherent 
modes the variational wavefunctions of Ref. [3], it is easy to check, by 
direct calculations, that the maximal values of $|(\varphi_m,\varphi_n)|$ 
for different $m\neq n$ are of order $0.1$. Moreover, even this approximate 
orthogonality is not compulsory in the frame of the quasiresonance picture 
employed in Ref. [3]. For deriving Eqs. (3), it is sufficient that $c_n(t)$ 
be considered as a slow function as compared to $\exp(-iE_nt/\hbar)$. Then, 
integrating Eq. (1), one can make use of the {\it exact orthonormality on 
average} of functions $\varphi_n$ in the sence of the equality 
$\lim_{\tau\rightarrow\infty}\frac{1}{\tau}\int_0^\tau\left [ \int
\varphi^*_m(\vec r,t)\varphi_n(\vec r,t)d\vec r\right ] dt=\delta_{mn}$.
Calculations of Ref. [3] show that $c_n(t)$ can, really, be treated as 
slow, compared to $\exp(-iE_nt/\hbar)$ since the variation rate of $c_n$ 
is of order $\alpha$ which is an order smaller than $E_n$, that is,
$|dc_n/dt|\ll E_n$.

We solved the system of nonlinear equations (3) numerically, carefully
analyzing the behaviour of the solutions for different parameters.
This behaviour turned out to be surprisingly rich exhibiting unexpected
critical effects. Analyzing Eqs. (3), it is convenient to make there a
scaling, measuring time in units of $\alpha^{-1}$ and introducing the
dimensionless parameters $b\equiv|\beta|/\alpha$ and $\delta\equiv\Delta
\omega/\alpha$. The dimensionless detuning is assumed to be always small,
$\delta\ll 1$. When the dimensionless transition amplitude $b$ is also small,
the fractional populations oscillate according to the sine-squared law.
When $b$ increases, the amplitude of oscillations also increases. The overall
behaviour continues to be normal unless we reach a critical value $b_c$, when
the dynamics of the system changes drastically. More generally, there exists
a critical line connecting the parameters $b$ and $\delta$, so that
$b_c\cong0.5-\delta$. In the vicinity of this line, the system dynamics
experiences sharp changes, when the parameters are varied just a little. We
illustrate this in Figs. 1--4, slightly varying the detuning and keeping
$b=0.4999$, so that we are close to the critical line. In Fig. 1, the
detuning is zero, and the oscillations of the fractional populations 
are yet normal. Shifting the detuning to $\delta=0.0001$ transforms the
picture to that in Fig. 2, where the top of $n_j(t)$ and the bottom of
$n_0(t)$ become flat, while the oscillation period is approximately doubled.
Changing the detuning to $\delta=0.0001001$ results in Fig. 3, where the
period is again doubled and there appear the upward cusps of $n_j(t)$ and
the downward cusps of $n_0(t)$. Increasing further the detuning to
$\delta=0.00011$ squeezes the oscillation period twice, as is shown in
Fig. 4. The same phenomena occur when we cross the critical line 
$b+\delta\cong 0.5$ at other values of parameters.

This unusual behaviour of the fractional populations is certainly due
to the nonlinearity of the evolution equations (3). Systems of nonlinear
differential equations, as is known, can possess qualitatively
different solutions for parameters differing by infinitesimally
small values. The transfer from one type of solutions to another type, in
the theory of dynamical systems, is termed bifurcation. At a bifurcation
point, dynamical system is structurally unstable. Bifurcations in
dynamical systems are somewhat analogous  to phase transitions and critical
phenomena in equilibrium statistical systems. To elucidate this analogy
for the present case, we have to consider the time-averaged behaviour of the
system, which can be done as follows. First, we need to define an 
effective Hamiltonian generating the evolution equations (3). To this 
end, we notice that the latter equations can be presented in the Hamiltonian 
form $idc_0/dt=\partial H_{eff}/\partial c_0^*,\; idc_j/dt=
\partial H_{eff}/\partial c_j^*$, with the effective Hamiltonian
\begin{equation}
\label{4}
H_{eff} = \alpha\; n_0n_j + \frac{1}{2}\left ( \beta\; e^{i\Delta\omega t}
c_0^*c_j + \beta^*\; e^{-i\Delta\omega t} c_j^* c_0\right ) \; .
\end{equation}
Then, using the averaging method, we solve the evolution equations (3)
finding 
$$
c_0 =\left [ \cos\frac{\Omega t}{2} +
i\; \frac{\alpha(n_0-n_j)-\Delta\omega}{\Omega}\;\sin\frac{\Omega t}{2}
\right ]\exp\left\{ -\;\frac{i}{2}(\alpha-\Delta\omega)\; t\right\}\; ,
$$
\begin{equation}
\label{5}
c_j=-\; i\;\frac{\beta^*}{\Omega}\;\sin\frac{\Omega t}{2}\;
\exp\left\{ -\; \frac{i}{2}(\alpha +\Delta\omega)\; t\right\} \; , \qquad
\Omega^2 =\left [ \alpha(n_0-n_j)-\Delta\omega\right ]^2 + |\beta|^2 \; .
\end{equation}
The usage of the averaging method here is accomplished in the same way as 
it has been done for solving nonlinear evolution equations describing 
superradiant spin relaxation in magnets [10,11], or nonadiabatic dynamics 
of atoms in magnetic traps [12,13]. Technical details can be found in the 
latter references. Employing the found solutions (5), with the normalization
$n_0+n_j=1$, we obtain for the effective Hamiltonian 
$H_{eff}=\alpha\; n_j^2 +n_j\Delta\omega$. Averaging the fractional 
populations over the explicitly entering time, we get the average 
population $\overline n_j=b/2\varepsilon$, where, as earlier, 
$b\equiv|\beta|^2/\alpha$ and $\varepsilon\equiv\overline\Omega/\alpha$ is 
the dimensionless average frequency, with $\overline\Omega$ given
in Eqs. (5), where $n$ is replaced by $\overline n_j$. This average 
frequency is defined by the equation
$\varepsilon^4(\varepsilon^2-b^2)=(\varepsilon^2-b^2-\varepsilon^2\delta)^2$.
Substituting the average population $\overline n_j$ in the effective
Hamiltonian, we obtain the average effective energy
$E_{eff} =(\alpha b^2/2\varepsilon^2)(b^2/2\varepsilon^2+\delta)$.
An order parameter for the averaged system can be defined as the difference
of the average populations, $\eta\equiv\overline n_0 -\overline n_j=
1-b^2/\varepsilon^2$. The capacity of the system to store the energy 
pumped in by the resonant field is characterized by the derivative 
$C_\beta=\partial E_{eff}/\partial|\beta|$ of the average energy 
with respect to $|\beta|$. And the absolute value of the derivative of the
order parameter with respect to $\delta$ defines a kind of susceptibility
$\chi_\delta=|\partial\eta/\partial\delta|$. Note that $C_\beta$ plays 
the role of heat capacity, because of which it can be called the pumping 
capacity.

The solution for $\varepsilon$ shows that the average frequency changes from
$\varepsilon\cong 1-2\delta$ at $b=0$ to $\varepsilon\cong\sqrt{0.5-\delta}$
at $b=b_c\equiv0.5-\delta$. After this, the frequency diminishes by a jump
to $\varepsilon\cong b$ for $b>b_c$. This implies that at the point $b_c$ the
average oscillation period increases by a jump almost twice. The order
parameter diminishes from $\eta=1$, when there is no resonance modulation,
that is $b=0$, to $\eta_c\cong 0.5+\delta$ at $b=b_c$, after which the order
parameter becomes approximately zero for $b>b_c$. The pumping capacity grows
from zero at $b=0$, becoming divergent at $b=b_c$. The same concerns the
susceptibility. All this suggests that $b_c$ is a critical point and the
relation $b_c+\delta=0.5$ defines the critical line, which is in agreement
with our numerical solution of the evolution equations (3). The asymptotic
behaviour of the characteristic quantities in the vicinity of the
critical line defines the corresponding critical indices. Thus, for the small
relative deviation $\tau\equiv|b-b_c|/b_c\rightarrow 0$ from the critical
point $b_c$, we have
$\eta-\eta_c\simeq \frac{\sqrt{2}}{2}\; ( 1 - 2\delta)\tau^{1/2},\;
C_\beta\simeq \; \frac{\sqrt{2}}{8}\; \tau^{-1/2},\;
\chi_\delta \simeq\; \frac{1}{\sqrt{2}}\;\tau^{-1/2}$. Hence, the related 
critical indices are $1/2$ for all characteristics. It is interesting 
that the critical indices of $\eta,\; C_\beta$, and $\chi_\delta$
satisfy the known scaling relation: $ind(C_\beta)+2\; ind(\eta)+ 
ind(\chi_\delta)=2$, where $ind$ is the evident abbreviation for index.

If, after exciting a coherent mode, one switches off the pumping, then 
the following dynamics is to be considered separately, taking into 
account stability conditions as well as the lifetime of the condensate
as such, which is defined by depolarizing collisions [14]. Finally, let 
us analyse the optimal conditions when the considered effects could be 
achieved experimentally. To reach the bifurcation line one needs to invoke 
the transition amplitude $\beta\approx \alpha/2$. At the same time, the 
interaction parameter $\alpha$, at least in the strong-coupling limit [3], 
can be of the order of the transition frequency $\omega$. When the magnitude 
of the transition amplitude $\beta$ becomes close to the transition frequency 
$\omega$, the effect of power broadening is expected. This effect is well 
known in optics [15], where the role of $\beta$ is played by the Rabi 
frequency. In the regime of power broadening, one would expect the ground 
state to  be coupled to more than one excited mode, so that the 
quasi-resonant two-mode picture could become not a good approximation. 
This picture remains a good approximation provided that the probability of 
nonresonant  excitation of neighboring levels is small. The probability 
of an induced transition between levels $i$ and $j$, due to a monochromatic 
field with frequency $\omega$, can be estimated [15] as 
$P_{ij}\approx\beta_{ij}^2/2[(\omega-\omega_{ij})^2+\beta_{ij}^2]$, where 
$\beta_{ij}$ and $\omega_{ij}$ are the related transition amplitude and 
frequency,  respectively. To estimate $P_{ij}$, we may use the calculations 
of Ref. [3], slightly modifying them to take into account the cylindrical 
symmetry of a trap with radial and axial frequencies $\omega_r$ and 
$\omega_z$, respectively. Dealing with a cylindrical trap gives an 
additional explicit parameter $\nu\equiv\omega_z/\omega_r$, hence, more 
possibilities for varying conditions. In the case of nonresonant 
excitation, when $\omega-\omega_{ij}\sim\omega$, the magnitude of 
$P_{ij}$ essentially depends on the value of the transition amplitude 
$\beta_{ij}$. The maximal value of the latter, occurring when approaching
the bifurcation line, is close to $\alpha/2$. The value of $\alpha$ 
depends on the parameter $g\nu$, with $g\equiv 4\pi a_sN/l_r$, where 
$a_s$ is a scattering length, $N$ is the number of atoms, and 
$l_r\equiv\sqrt{\hbar/m_0\omega_r}$. If $g\nu$ is not large, then 
$\alpha\ll\omega$. Thence the maximal $\beta_{ij}\approx\alpha/2\ll\omega$ 
and $P_{ij}\ll 1$. Therefore the nonresonant excitation can be neglected, 
and the two-level picture is a good approximation. When $g\nu\gg 1$, then
$\alpha\sim\omega$, so that the maximal $\beta_{ij}\approx\omega/2$, from 
where $P_{ij}\sim 0.1$. This tells that, although power broadening can 
influence quantitative results, the two-level picture still serves as a 
reasonable first approximation. To reduce the influence of power broadening, 
one can proceed as follows. In the case of an almost spherical trap, with 
$\nu\approx 1$, the number of atoms is to be restricted in order not to 
make $g$ too high. When the number of atoms is large, so that $g\gg 1$, this 
can be compensated by taking a long cigar-shape trap, with $\nu\ll 1$, which 
would result in $g\nu\sim 1$. In the present-day experiments there is such a 
wide variety of possibilities for changing characteristic parameters [1,2] by 
choosing an appropriate shape of a trap, by changing the number or the type 
of trapped atoms, by varying the scattering length using Feshbach resonances, 
and so on, that it looks rather feasible to observe the critical 
phenomena described.

\newpage

\begin{center}
{\bf Figure captions}
\end{center}

{\bf Fig. 1.} The time dependence of the fractional populations $n_0(t)$ and
$n_j(t)$ for $b=0.4999$ and $\delta=0$. Here and in all following pictures
the dashed line corresponds to the ground-state population $n_0(t)$ and the
solid line, to the excited-level population $n_j(t)$.

\vskip 5mm

{\bf Fig. 2.} Flattening of the fractional populations, with their
oscillation period being doubled, at $b=0.4999$ and $\delta=0.0001$.

\vskip 5mm

{\bf Fig. 3.} The appearance of the upward cusps of $n_j(t)$ and of the
downward cusps of $n_0(t)$ for $b=0.4999$ and $\delta=0.0001001$.

\vskip 5mm

{\bf Fig. 4.} Fractional populations versus time for $b=0.4999$ and
$\delta=0.00011$.

\end{document}